# Spontaneous Chirality Flipping in an Orthogonal Spin-Charge Ordered Topological Magnet

H. Miao[1,*,#], J. Bouaziz[2,*,#], G. Fabbris[3,*], W. R. Meier[4], F. Z. Yang[1], H. X. Li[1,5], C. Nelson[6], E. Vescovo[6], S. Zhang[7], A. Christianson[1], H. N. Lee[1], Y. Zhang[4,8], C. D. Batista[4,9], S. Blügel[2]

[1]*Materials Science and Technology Division, Oak Ridge National Laboratory, Oak Ridge, Tennessee 37831, USA*

[2]*Peter Grünberg Institut and Institute for Advanced Simulation, Forschungszentrum Jülich &JARA, D-52425 Jülich, Germany*

[3]*Advanced Photon Source, Argonne National Laboratory, Argonne, Illinois 60439, USA*

[4]*Department of Physics and Astronomy, The University of Tennessee, Knoxville, Tennessee, 37996, USA*

[5]*Advanced Materials Thrust, The Hong Kong University of Science and Technology (Guangzhou), Guangzhou, China*

[6]*National Synchrotron Light Source II, Brookhaven National Laboratory, Upton, New York 11973, USA*

[7]*Max-Planck-Institut fur Physik komplexer Systeme, Nothnitzer Straße 38, 01187 Dresden, Germany*

[8]*Min H. Kao Department of Electrical Engineering and Computer Science, University of Tennessee, Knoxville, Tennessee 37996, USA*

[9]*Quantum Condensed Matter Division and Shull-Wollan Center, Oak Ridge National Laboratory, Oak Ridge, Tennessee 37831, USA*

*These authors are contributed equally.
#Correspondence should be addressed to miaoh@ornl.gov; j.bouaziz@fz-juelich.de.

**The asymmetric distribution of chiral objects with opposite chirality is of great fundamental interests ranging from molecular biology to particle physics[1]. In quantum materials, chiral states can build on inversion-symmetry-breaking lattice structures or emerge from spontaneous magnetic ordering induced by competing interactions. Although the handedness of a chiral state can be changed through external fields[2,3], a spontaneous chirality flipping has yet to be discovered. In this letter, we present experimental evidence of chirality flipping via changing temperature in a topological magnet $EuAl_4$, which features orthogonal spin and charge density waves (SDW/CDW). Using circular dichroism of Bragg peaks in the resonant magnetic x-ray scattering, we find that the chirality of the helical SDW flips through a first order phase transition with modified SDW wavelength. Intriguingly, we observe that the CDW couples strongly with the SDW and displays a rare commensurate-to-incommensurate transition at the chirality flipping temperature. Combining with first**

**principles calculations and angle resolved photoemission spectroscopy, our results support a Fermi surface origin of the helical SDW with intertwined spin, charge, and lattice degrees of freedom in EuAl$_4$. Our results reveal an unprecedented spontaneous chirality flipping and lays the groundwork for a new functional manipulation of chirality through momentum dependent spin-charge-lattice interactions.**

Chirality, a geometrical concept that distinguishes an object from its mirror image, has been proposed for over three decades as a potential mechanism for novel quantum states including spontaneous quantum Hall liquids[4], chiral spin liquids[5], and magnetic skyrmions[6,7]. Recently, chirality has experienced a revival in the context of correlated and geometrically frustrated electronic systems[8-15]. In these settings, chiral spin, charge, orbital, and pairing fields become strongly coupled, giving rise to intertwined orders[13] and long-range entangled quasiparticles[14,15]. In magnetic systems, the 1D and 2D chiral spin textures, as respectively shown in Fig. 1**b** and **c**, have been widely observed in non-centrosymmetric lattices[2,16]. In these systems, the chirality or handedness, $\chi=\pm1$, is usually transmitted to the spin system via the relativistic Dzyaloshinskii–Moriya (DM) interaction induced by spin-orbit coupling. Chiral spin orders can also spontaneously emerge in centrosymmetric materials[2,16-19], where competing magnetic interactions, such as the Ruderman–Kittel–Kasuya–Yosida (RKKY), can yield equally populated $\chi=1$ and -1 states. Experimentally, chirality manipulation has been achieved by applying external fields[2,3]. An outstanding question that remains is if the sign of chirality can be controlled by other means.

In this letter, we uncover an unprecedented spontaneous chirality flipping in EuAl$_4$. EuAl$_4$ hosts nanometric skyrmions and the topological Hall effect under a magnetic field along the c-axis[20-22]. At zero external magnetic field, EuAl$_4$ exhibits a tetragonal structure with *I4/mmm* symmetry (centrosymmetric space group No. 139) at room temperature. Figure 1**d** summarizes the zero magnetic field symmetry breaking orders of EuAl$_4$ (see also Supplementary Figure S1 and Table S1). Below $T_{CDW}$=140 K, an incommensurate charge density wave (ICDW) develops along the crystalline c-axis with $\boldsymbol{Q}_{CDW}$=(0, 0, 0.183) in reciprocal lattice unit (r.l.u.). The $\boldsymbol{Q}_{CDW}$ gradually shifts to the commensurate position at (0, 0, 1/6) via decreasing temperature, which is typical for ICDW systems due to the lattice commensurate energy[23]. At $T_{SDW}^{(1)} = 15.4$ K, the system breaks the time-reversal symmetry, $\mathcal{T}$, by forming a double-$\boldsymbol{Q}$ spin density wave (SDW) along the [110] and [1-10] directions. Below $T_{SDW}^{(2)} = 13.3$ K, the spin moment of the double-$Q$ SDW gains a finite c-axis component, leading to new interference magnetic peaks along the [100] and [010] direction. The fourfold rotational symmetry, $C_4$, in the ab-plane is broken at $T_{SDW}^{(3)}$=12.3 K, resulting in a stripe helical SDW[20-21]. Interestingly, although the helical SDW persists down to the lowest temperature at zero magnetic field, an additional first order phase transition sets in at $T_\chi$=10.1 K.

Here we use the circular dichroism (CD) of Bragg peaks in the resonant magnetic x-ray scattering (XRMS) to demonstrate that the first order phase transition at $T_\chi$ leads to a spontaneous chirality flipping. CD-XRMS is a direct experimental probe of chiral electronic orders[24-28]. Under the resonance condition, where the incident photon energy, $\omega$, matches the energy differences between occupied and unoccupied atomic energy levels, the x-ray scattering amplitude from site $n$ can be written as[24] (see Methods):

$$f_n(\omega) = (\hat{\epsilon}' \cdot \hat{\epsilon})f_0(\omega) - i(\hat{\epsilon}' \times \hat{\epsilon}) \cdot \widehat{M_n}f_1(\omega) \quad (1)$$

where $\hat{\epsilon}$ and $\hat{\epsilon}'$ are the polarization vectors of the incident and scattering x-rays, respectively, and $\widehat{M_n}$ is the magnetic moment of site $n$. $f_0(\omega)$ is the anomalous charge scattering form factor that can be added to the Thomson scattering. $f_1(\omega)$ is the linear magnetic scattering form factor. For the experimental geometry shown in Fig. 2**a**, the CD of the helical SDW can be formulated as[28]:

$$I(Q)^{CR} - I(Q)^{CL} = (\tau\chi)\mathcal{D}^{yz} \quad (2)$$

where $\tau=sign((\boldsymbol{Q}_{SDW} \cdot \vec{x})/|\boldsymbol{Q}_{SDW} \cdot \vec{x}|)$, $\vec{x}$ is the unit vector along the x-direction as shown in Fig. 2**a**. $\mathcal{D}^{yz}$ is a function of magnetic moment in the (y-z) plane (Fig. 2**a**) and is independent of $\tau$ and $\chi$. $I(Q)^{CR}$ and $I(Q)^{CL}$ represent x-ray intensity obtained under circular right (CR) and circular left (CL) incident photon energy, respectively. Following Eq. (2), the CD of the $\chi$=1 helical SDW (Fig. 1**b**) is positive for a propagation vector $\boldsymbol{Q}_{SDW}$ and negative for -$\boldsymbol{Q}_{SDW}$. An achiral SDW, such as the double-$Q$ SDW above $T_{SDW}^{(3)}$ will, therefore, yield zero CD. To probe the magnetic chirality of EuAl$_4$, the photon energy is tuned to the Eu $L_3$-edge (2$p$-5$d$). Figure 2**a** shows the x-ray fluorescence scan at $T$=5 K. The single peak at $\omega_{res}$=6.977 keV confirms the Eu$^{2+}$ electronic configuration in EuAl$_4$[21]. The energy-scan at fixed $\boldsymbol{Q}_{SDW}$=(0.19, 0, 4) at 5 K show giant magnetic resonance at $\omega_{res}$, confirming its magnetic origin.

In Fig. 2**b** and 2**c**, we first show the CD of a structural Bragg peak (0, 0, 4) at *T*= 5K. Yellow and cyan curves represent $I(Q)^{CR}$ and $I(Q)^{CL}$, respectively. The asymmetry of the CD, $F(Q) = I(Q)^{CR} - I(Q)^{CL}/I(Q)^{CR} + I(Q)^{CL}$, is shown in Fig. 2**c**. As expected, the (0, 0, 4) structural Bragg peak shows *F(Q)*=0. Note the large noise away from the Bragg condition is due to the nearly zero scattering intensity, proving the high sample quality. We then move to the helical SDW at T=9 K<$T_\chi$. As shown in Fig. 2**d** and 2**e**, we observe giant CD at both $Q_{SDW}$ and -$Q_{SDW}$ with $F(Q_{SDW})$=40% and $F(-Q_{SDW})$=-90%. This observation proves that the helical SDW below $T_\chi$ is chiral with χ=1. The large *F* indicates that the entire photon illuminated SDW (on the order of 50 μm × 50 μm) has the same χ and hence macroscopically breaks the symmetric chiral distribution. We then move to $T_\chi$ <T=11 K<$T_{SDW}^{(3)}$. Remarkably, as shown in Fig. 2**f** and 2**g**, the CD changes sign with $F(Q_{SDW})$ =-40% and $F(-Q_{SDW})$ =90%. This observation establishes a spontaneous chirality flipping from χ=1 to χ=-1. The comparably large $F(Q)$ below and above $T_{SDW}^{(4)}$ further suggests that the chirality flipping is also realized on a macroscopic length scale. The chiral density is back to nearly zero upon warming up above $T_{SDW}^{(3)}$. The giant asymmetric chiral distribution and spontaneous chirality flipping between the helical SDW states constitute the main experimental results of this work.

The spontaneous chirality flipping raises questions concerning its microscopic origin. Due to the coexistence of CDW and SDW, we first determine the complex spin-charge correlations by tracing the temperature dependent evolution of CDW and SDW wavevectors below 20 K. The scanning trajectories in the reciprocal space are shown in the inset of Fig. 3**a-c**. Figure 3**d** summarizes the extracted $Q_{DW}$ (in r.l.u.). As shown in Fig. 3**a**, the double-*Q* SDW first emerges below $T_{SDW}^{(1)}$ along the [110] and [1-10] directions and is smoothly connected with the spin canted double-*Q* phase. In the chiral SDW phase below $T_{SDW}^{(3)}$, $Q_{SDW}$ increases monotonically along the [100] and [010] direction and displays a discontinuous leap at the chirality flipping transition. For the CDW, the $Q_{CDW}$ first shows a rare commensurate-incommensurate transition in the temperature range [$T_{SDW}^{(3)}$, 20K]≪$T_{CDW}$. The $Q_{CDW}$ then jumps back to the commensurate value and remains *T*-independent in the χ=-1 SDW phase ([$T_\chi$, $T_{SDW}^{(3)}$]). Finally, in the χ=1 SDW phase (T<$T_\chi$), the $Q_{CDW}$ once again becomes incommensurate. This complex temperature-dependent evolution of the CDW and the SDW is characteristic of intertwined spin, charge, and lattice degrees of freedom.

The incommensurability of both $Q_{SDW}$ and $Q_{CDW}$ and the presence of large itinerant carriers in EuAl$_4$ indicate a Fermi surface effect. Figure 3**e** and **f** show the calculated 3D Fermi surface and Eu-Eu magnetic interaction, $J(\mathbf{q})$, of EuAl$_4$ in the tetragonal phase (see Supplementary Figure S2 for the electronic structure determined by angle-resolved photoemission spectroscopy). The highest value of $J(\mathbf{q})$ determines the Néel temperature and the wavevector $\mathbf{q}_p$ of the helical SDW state. As shown in Fig. 3**f**, $J(\mathbf{q}_p)$ along the [100] direction features a typical paramagnetic spin susceptibility of a metal with a sharp and significant finite-$q$ peak at $q_p$=0.19 r.l.u., consistent with experimental data at 5 K. The estimated magnetic transition temperature, $T_{SDW}^{cal} \propto J(\mathbf{q}_p)/3k_B \sim 14.8$ K ($k_B$ being the Boltzman constant), is also in agreement with experimental observation (see Supplementary Figure S3-S6 for the effects of Coulomb interactions, electron temperature and magnetoelastic coupling on $J(\mathbf{q})$). These findings provide strong numerical evidence for a Fermi surface driven helical SDW in EuAl$_4$. Interestingly, the $Q_{CDW}$ matches the calculated charge susceptivity peak along the Γ-Z direction[29]. Although the primary driving force of the CDW in EuAl$_4$ remains to be determined, the presence of nested Fermi surface is usually helpful to select the $Q_{CDW}$ by forming a CDW gap near the Fermi level[30,31].

While the intertwined spin, charge, and lattice degrees of freedom are established in EuAl$_4$, the microscopic origin of the giant asymmetric chiral distribution and spontaneous chirality flipping calls for further studies. The CD of the SDW is robust under temperature cycling above both the achiral double-Q phase and C4 symmetry-breaking (see Supplementary Figures S7 and S8), suggesting hidden chiral interactions above 20 K. Due to the intertwined nature of SDW and CDW, it is tempting to associate the chiral interaction with a chiral CDW. Encouragingly, the CDW in EuAl$_4$ is found to be transverse[32], where the CDW driven lattice distortions and soft phonon modes are perpendicular to the CDW propagation vector[29]. Assuming a linear transverse CDW along the propagation direction, the C4 rotational symmetry is expected to be broken below $T_{CDW}$. However, the C4 symmetry-breaking in the ab-plane is observed only at $T_{SDW}^{(3)}$=12.3 K<< $T_{CDW}$=140 K[20,21]. These observations, therefore, support a chiral CDW in EuAl$_4$. It is highly interesting to point out that the CDW related "nesting" vector connects the topological semi-Dirac bands[33]. Similar type of band structure has been studied in 3D-quantum Hall systems, where electron-electron scattering between the Dirac bands involves chiral lattice excitations[34,35]. The role of chiral phonons in EuAl$_4$ is therefore an interesting open question.

Finally, we discuss the origin of chirality flipping. Since chirality is a discontinuous physical quantity, the chirality flipping requires a first order phase transition that is consistent with the jump of $Q_{SDW}$ at $T_\chi$. The absence of hysteresis observed in this study and previous works[20,21] suggests that the transition is weak first order[31,36]. Since the itinerant electrons play a pivotal role for both CDW and SDW in EuAl$_4$, the chiral magnetic interaction is also likely momentum dependent (See Supplementary Note 1 and 2 for a simplified model). Depending on $Q$, the chiral interaction energetically favors $\chi=1$ or -1 SDW. Indeed, as we show in Fig. 3**a** and 3**d**, the chirality flipping accompanies with a significant jump of $Q_{SDW}$ but minor lattice and CDW modifications.

In summary, we discovered a spontaneous chirality flipping in an orthogonal spin-charge ordered itinerant magnet. Our results highlight EuAl$_4$ and associated materials as a rare platform for emergent chiral interactions and open a new avenue for chiral manipulations through intertwined orders.

**Methods:**

**Sample Growth:**

$EuAl_4$ crystals were grown from a high-temperature aluminum-rich melt[21,33]. Eu pieces (Ames Laboratory, Materials Preparation Center 99.99+%) and Al shot (Alfa Aesar 99.999%) totaling 2.5g were loaded into one side of a 2-mL alumina Canfield Crucible Set. The crucible set was sealed under 1/3 atm argon in a fused silica ampoule.

The ampoule assembly was placed in a box furnace and heated to 900 °C over 6 h (150 °C/h) and held for 12 h to melt and homogenize the metals. Crystals were precipitated from the melt during a slow cool to 700 °C over 100 h (−2 °C/h). To liberate the crystals from the remaining liquid, the hot ampoule was removed from the furnace, inverted into a centrifuge, and spun.

**XRMS:**

Resonant magnetic x-ray-scattering measurements were performed at the integrated *In situ* and resonant hard x-ray studies (4-ID) beam line of National Synchrotron Light Source II (NSLS-II). The photon energy, which is selected by a cryogenically cooled Si(111) double-crystal monochromator is 6.977 keV. The sample is mounted in a closed-cycle displex cryostat in a vertical scattering geometry, and the magnetic σ-π scattering channel is measured using an Al(222) polarization analyzer and silicon drift detector.

**CD-XRMS:**

The CD-XRMS were performed at the 4-ID-D beamline of the Advanced Photon Source (APS), Argonne National Laboratory (ANL). The photon energy was tuned to the Eu $L_3$ resonance (6.977 keV) using a double-crystal Si (111) monochromator. Circularly polarized x-rays were generated using a 180 μm thick diamond (111) phase plate[34], focused to 200x100 μm² full-width-half-maximum (FWHM) using a toroidal mirror, and further reduced to 50x50 μm² FWHM with slits. Temperature was controlled using a He closed-cycle cryostat. Diffraction was measured in reflection from the sample [001] surface using vertical scattering geometry and an energy dispersive silicon drift detector (approximately 0.15 keV energy resolution).

**ARPES:**

The ARPES experiments were performed on single crystals $EuAl_4$. The samples are cleaved *in situ* in a vaccum better than 3×10⁻¹¹ torr. The experiment was performed at beam line 21-ID-1 at the NSLS-II. The measurements are taken with synchrotron light source and a Scienta-Omicron DA30 electron analyzer. The total energy resolution of the ARPES measurement is approximately 15 meV. The sample stage is maintained at T=20 K throughout the experiment.

**DFT:**

The first principles simulations are performed using the all-electron full-potential Korringa-Kohn-Rostoker (KKR) Green function method[37] in the scalar relativistic approximations. The inclusion of the spin orbit coupling self consistently does not alter the magnetic interactions or Fermi surface. The $Eu^{2+}$ and its 4f-electrons are treated using DFT+U approach[38] with value of U=2 eV. The calculations are done using an angular momentum cutoff of lmax= 4 in the orbital expansion of the Green function. The self-consistent energy contour takes into 58 energy points with a Brillouin zone k-mesh of 30x30x30. The magnetic interactions are between the Eu magnetic atoms are computed in real space using the infinitesimal rotation method[39], then Fourier transformed into reciprocal space with a cutoff of 10 lattice constants to take account accurately the long RKKY interactions. The Fermi surface is imaged by computing q-dependent density of states[40].

**XRMS cross-section:**

Under the electric dipole approximation, the resonant scattering process involves transitions between the core state $|\zeta_v\rangle$ with energy $E_v$ and an unoccupied state $|\psi_\eta\rangle$ with energy $E_\eta$ in both absorption and emission channel. The scattering amplitude can be written as:

$$f_{res}(\omega) = \sum_{ij} \hat{\epsilon}'_i \hat{\epsilon}_j \sum_\eta \frac{\langle \zeta_v|R_i|\psi_\eta\rangle\langle\psi_\eta|R_j|\zeta_v\rangle}{\omega-(E_\eta-E_v)+i\Gamma} = \sum_{ij} \hat{\epsilon}'_i \hat{\epsilon}_j T_{ij} \quad (M1)$$

where $\hat{\epsilon}$ and $\hat{\epsilon}'$ are the polarization vectors of the incident and scattering x-rays, respectively. $R$ is the position operator. When the resonant atom has a parity-even magnetic moment $\widehat{M_n}$ at site $n$, and assuming cylindrical symmetry, the magnetic scattering takes the form:[24]

$$f_n(\omega) = (\hat{\epsilon}' \cdot \hat{\epsilon})f_0(\omega) - i(\hat{\epsilon}' \times \hat{\epsilon}) \cdot \widehat{M_n} f_1(\omega) + (\hat{\epsilon}' \cdot \widehat{M_n})(\hat{\epsilon} \cdot \widehat{M_n}) f_2(\omega) \quad (M2)$$

Since $f_2(\omega)$ is usually much smaller than $f_1(\omega)$, the magnetic scattering is dominated by the second term of Eq. (M2). For the 1D helical SDW, the CD has been derived respectively for the tilting and rocking geometries shown in Fig. 2a[28]. For the chiral Bloch type SDW:

$$I(Q)^{CR} - I(Q)^{CL} = (\tau\chi)\mathcal{D}^{yz} \quad \text{in the rocking geometry,} \quad (M3)$$

For the achiral Néel type SDW:

$$I(Q)^{CR} - I(Q)^{CL} = \chi \mathcal{B}^{yz} \quad \text{in the tilting geometry,} \quad (M4)$$

Note that for the Néel type SDW, the CD doesn't change sign from the positive to negative propagation vector. Our experiment was performed in the rocking geometry that shows sign change for the chiral Bloch type SDW.

**Non-resonant XRMS cross-section:**

For non-resonant scattering, the x-ray scattering amplitude can be written as:

$$f_{non}(\omega) \propto \sum_j <0|e^{i\vec{q}\cdot\vec{r}_j}|> (\vec{\epsilon}'^* \cdot \vec{\epsilon}) - i\frac{\hbar\omega}{mc^2}\left[\frac{mc}{e\hbar} <0|\hat{q}\times(\vec{M}_L(\vec{q})\times\hat{q})|0> \cdot \vec{P}_L + \frac{mc}{e\hbar} <0|\left(\vec{M}_S(\vec{q})\right)|0> \cdot \vec{P}_S\right]$$
$$(M5)$$

where $\vec{M}_L$ and $\vec{M}_S$ are Fourier transform of orbital and spin moment density, respectively. Here,

$$\vec{P}_L = 4\sin^2\theta(\vec{\epsilon}'^* \times \vec{\epsilon}) \quad (M6)$$

$$\vec{P}_S = \vec{\epsilon} \times \vec{\epsilon}' + (\hat{k}_f \times \vec{\epsilon}'^*)(\hat{k}_f \cdot \vec{\epsilon}) - (\hat{k}_i \times \vec{\epsilon})(\hat{k}_i \cdot \vec{\epsilon}'^*) - (\hat{k}_f \times \vec{\epsilon}'^*) \times (\hat{k}_i \times \vec{\epsilon}) \quad (M7)$$

Equations (M5)-(M7) shows that the non-resonant x-ray scattering can also probe spin and orbital magnetic moment and display CD. However, the factor $\frac{\hbar\omega}{mc^2}$ in Eq. (R1) at $\hbar\omega \sim 10$ keV is on the order of $\sim 10^{-4}$. Therefore, for the non-resonant x-ray scattering, the cross-section related to Eq. (M6) and (M7) is extremely small, typically 10~30 counts/s for magnetic materials.

In our study, we not simply observed CD, but the asymmetry of the CD that is described in Eq. (M4). Furthermore, $F(Q) = I(Q)^{CR} - I(Q)^{CL}/I(Q)^{CR} + I(Q)^{CL}$ ~90%, is very large, excluding contributions from Eqs. (M6) and (M7) as the origin of chirality flipping.

**Acknowledgements:** We thank Matthew Brahlek, Miao-Fang Chi, Xi Dai, Satoshi Okamoto, Andrew May, Brian Sales, Jiaqiang Yan, Gabriel Kotliar for stimulating discussions. This research was supported by the U.S. Department of Energy, Office of Science, Basic Energy Sciences, Materials Sciences and Engineering Division (x-ray and ARPES measurement). CD-XRMS used resources (beamline 4ID) of the Advanced Photon Source, a U.S. DOE Office of Science User Facility operated for the DOE Office of Science by Argonne National Laboratory under Contract No. DE-AC02-06CH11357. ARPES and XRMS measurements used resources at 21-ID-1 and 4-IDbeamlines of the National Synchrotron Light Source II, a US Department of Energy Office of Science User Facility operated for the DOE Office of Science by Brookhaven National Laboratory under contract no. DE-SC0012704.

**Figure 1: Emergent chiral magnetic orders and the zero-field phase diagram of EuAl$_4$. a**, Spontaneous chiral symmetry breaking yields degenerated $\chi=1$ and $\chi=-1$ states. External field or intertwined orders can lift the degeneracy. **b**, schematics of 1-dimensional (1D) and 2D chiral spin textures, helical SDW (left) and Bloch-type Skyrmion (right). **c**, possible microscopic mechanisms that drive chiral magnetic states. **Left**: relativistic DM-interaction, **D**, in non-centrosymmetric lattice determines the sign of $\chi$ (See Supplementary Note 1). The wavelength of the spin order, $\lambda$, is proportional to the relative energy scales of atomic exchange energy, **J**, and **D**. **Right:** quasi-nested Fermi surface in hexagonal and tetragonal structures give rise to frustrated RKKY interactions along **$Q_1$**, **$Q_2$**, **$Q_3$** that satisfy **$Q_1$**+**$Q_2$**+**$Q_3$** =0. The inset depicts possible Fermi surface topologies in the hexagonal and tetragonal lattices that can yield nearly degenerated **$Q_1$**, **$Q_2$**, **$Q_3$** orders with **$Q_1$**+**$Q_2$**+**$Q_3$** =0. The Fermi surface topology of EuAl$_4$ is similar to the one in the tetragonal lattice, where yellow and cyan represent electron and hole band, respectively (Supplementary Figure S4). **d**, phase diagram of EuAl$_4$ without external magnetic field. Orthogonal CDW and SDW superlattice peaks are marked in the 3D and 2D Brillouin zone, respectively. Chiral SDW emerges below $T_{SDW}^3$ and accompanies with $C_4$ symmetry breaking.

**Figure 2: Discovery of spontaneous chirality flipping in EuAl$_4$. a**, experimental geometry of CD-XRMS. The photon energy was tuned to Eu $L_3$-edge to probe the magnet order parameter. For a 1D chiral SDW, the CD is given by Eq. (2), where the sign of F(**Q**) changes from **Q**=**Q**$_{SDW}$ to **Q**=-**Q**$_{SDW}$ in the rocking scattering geometry. Here we define $\chi$=+1 if F(**Q**$_{SDW}$)>0. Fluorescence scan (bottom left) at T=5 K shows single peak at $\omega_{res}$=6.977 keV, confirming Eu$^{2+}$ configuration. Magnetic resonance scan (bottom right) at **Q**$_{SDW}$=(0.169, 0, 4). The strong resonant enhancement confirms its magnetic origin. **b-i**, CD of the structural and magnetic Bragg peaks. Yellow and cyan curves in **b**, **d**, **f**, and **h** represent CR and CL incident photon polarization, respectively. Red, green, and blue curves in **c**, **e**, **g**, and **i** represent positive, zero and negative F(**Q**). We note that due to the finite H component and narrow width of magnetic peaks, horizontal axis from H=-0.17 to 0.17 was not shown in **e**, **g**, and **i**. Giant CD is observed below $T_{SDW}^3$=12.3 K. The sign change of the CD shown in **e** and **g** establishes the chirality flipping across $T_\chi$.

**Figure 3: Intertwined SDW and CDW with orthogonal wavevectors. a** and **b,** *T*-dependent H-scan and HK-scan near the SDW wave vectors. **c**, *T*-dependent L-scan near the CDW wave vector.

The scanning trajectories are shown in each panel. Dashed lines indicate the magnetic transition temperatures. **d**, extracted *T*-dependent CDW and SDW wavevectors. Blue and purple squares are corresponding to |Q$_{SDW}$| in the unit of $\frac{2\pi}{a_0}$ and $\frac{2\pi}{a_0}\sqrt{2}$, respectively. In the spin canted double-Q phase between $T_{SDW}^{(2)}$ and $T_{SDW}^{(3)}$, the c-axis spin component yields superlattice peaks along [100] and [010] directions which values are $\sqrt{2}$ times of the principal peak values along [1-10] and [110] directions. Yellow circles represent |Q$_{CDW}$| in the unit of $\frac{2\pi}{c_0}$. $a_0$ and $c_0$ are lattice constants in the tetragonal unit cell. **e**, DFT calculated electronic structure in the tetragonal phase. Green arrows indicate the in-plane "nesting vector" that favors helical SDW. **f**, Calculated RKKY interactions along [100] (black) and [110] (red) directions. The prominent peak at $q_p$=0.19 r.l.u. is consistent with experimentally observed SDW at 5 K.